\documentclass{ws-p8-50x6-00}

\def\lsim{\mathrel{\vcenter{\hbox{$<$}\nointerlineskip\hbox{$\sim$}}}}

\begin{document}

\title{Singly produced new physics particles in future colliders}

\author{Katri Huitu}

\address{Helsinki Institute of Physics, P.O.Box 64,
FIN-00014 University of Helsinki, Finland\\
E-mail: katri.huitu@helsinki.fi}


\maketitle

\abstracts{
Single production at hadron and photon colliders
of gluinos and sneutrinos of supersymmetric
models, as well as radions of the Randall-Sundrum model, is
discussed.
In the case of supersymmetry, R-parity breaking is needed in
order to produce single susy particles.
Resonant production of radions is considered at $\gamma\gamma$
collider.
}

The main motivation for considering single production of
previously unobserved particles is the lower production threshold
when compared to the pair production.
Also, there may be enhancement in the cross section, if the
new particle can be produced as an $s$-channel resonance.

The most interesting scenarios for new physics from several previous
years are supersymmetry and models inspired by the extra dimensions.
We will consider here R-parity breaking supersymmetric model and
Randall-Sundrum model of extra dimensions \cite{rs}.
The talk is based on the references \cite{chy1,chy2,chky}.

\section{R-parity violation}

The R-parity, $R_p=(-1)^{3B+L+2S}$ ($L$ is lepton number,
$B$ is baryon number, and $S$ is spin), is often used in the
minimal supersymmetric standard model (MSSM) to
disallow terms, which would lead to fast proton decay.
It is well-known that it would be enough to have either lepton
or baryon number conservation instead of the whole R-parity.
Adding a subset of the possible R-parity violating ($R\!\!\!\!/_p$)
terms,
\bea
W_{R\!\!\!/_p}&=&\lambda_{[ij]k}L_iL_j\bar{E}_k+
\lambda^{'}_{ijk}L_iQ_j\bar{D}_k
+\lambda^{''}_{i[jk]}\bar{U}_i\bar{D}_j\bar{D}_k+\epsilon_iL_iH_u,
\nonumber
\eea
makes it possible to produce superparticles singly, and the lightest
supersymmetric particle (LSP) will decay.
The violation of R-parity has several interesting phenomenological
consequences, which are not allowed in the MSSM, e.g. neutrinos become
massive and they mix.

The R-parity violating couplings are bounded by experimental limits.
A list of the bounds is reported e.g. in \cite{add}.

\subsection{The process $pp(p\bar p)\rightarrow (t+\bar t)\tilde g$}

We consider two different schemes.
In the first one we assume universal boundary conditions at the
GUT scale and radiative symmetry breaking.
At one-loop we then have the relation
$M_1/g_1^2=M_2/g_2^2=M_3/g_3^2$, which tells us that the gluinos
are the heaviest gauginos at the weak scale.
The other scenario, which is considered, has a heavy gluino as LSP
\cite{raby}.
There have been worries \cite{fop} that this kind
of gluino would produce too many energetic neutrinos when annihilating
in the sun, but this problem is solved, if R-parity is broken.

In the process also top quarks are produced singly.
When compared to the other R-parity breaking single top production
processes, which have been studied,
the advantage of our calculation is that we have ${\lambda''}^2$,
not ${\lambda''}^4$.

Detection possibilities are determined by
the decay modes of the produced particles.
For top, only the decay $t\rightarrow bW\rightarrow bl\nu $ is
considered, since it is experimentally the easiest.
The gluino decay patterns are more complicated, see e.g. \cite{bartl},
and depend on the
gluino and other supersymmetric particle masses.
If gluino is the LSP or $\lambda''$ coupling
is large, the decay mode $\tilde g\rightarrow q_iq_jq_k$ becomes
important.
There are two $b$ or $t$ quarks in 60-70\% of the decays, and the
final state is
$t\tilde g\rightarrow 3b +nl+$missing energy ($n$=1,2,3).

Concerning the possible light gluino window around 25 GeV$\lsim
m_{\tilde g}\lsim$ 35 GeV, it seems that reanalysing Tevatron Run I
data would either close the window or improve the limits on couplings
and masses.
For heavier gluino, either LSP or not, LHC can put strict constraints on
the coupling constants or masses.

\subsection{The process $\gamma\gamma\rightarrow\tilde\nu$}

The advantage of the linear collider is its clean background when
compared to the hadron colliders.
In the $\gamma\gamma$ collision mode, a resonance can be probed over a
wide mass range, and couplings involving heavy flavours can be
introduced.
In addition to being loosely bounded, with an assumption of family
symmetry, one may expect the couplings involving heavy flavours to be
larger than those involving light flavours.

The productions of $\tilde\nu_2$ ($b$ quark or $\tau$ lepton in the
loop) and $\tilde\nu_3$ ($b$ quark in the loop) were studied.
If sneutrino and Higgs are of similar mass, the R-parity violating
flavour nondiagonal decays are essential for detection.
(Also more involved calculation for $\gamma\gamma\rightarrow\chi^\pm
l^\mp$ was performed, but it is dominated by the resonant sneutrino
production.)

Single sneutrino production $\gamma\gamma\rightarrow\tilde\nu
+f\bar{f}'$ with comparable cross section was studied in \cite{chry}.
The major advantage of this process is that the R-parity violating
sources may be distuinguished from the final state.

\section{Radions in the process $\gamma\gamma\rightarrow\Phi$}

A recently popular solution to the hierarchy problem is
provided by extra dimensions \cite{rs}.
Stabilization of the modulus leads to a physical light particle,
radion \cite{gw}.
The radion, $\Phi$, may be lighter than the lowest lying Kaluza-Klein
modes of graviton, and are thus the first particles, which are typical
for the model and possibly accessible in future collider experiments.

In general, radion can mix with Higgs with a mixing parameter $\xi$
\cite{grw}.
The peculiarity of the kinematical mixing is that the resulting
mixing matrix is nonunitary, and the matrix elements have
discontinuities.
Detection of lighter than 800 GeV radion seems possible for
$\sqrt{s}=1$ TeV, if the radion VEV is 1 TeV \cite{chky}.
If mixing is large, also a light radion with VEV of 10 TeV may be
possible to detect.
The different decay patterns of radion and Higgs make it
possible to distuinghuish between these two particles.

\section*{Acknowledgments}
The author thanks the organizers of SUSY'01 in Dubna for
an inspiring and well organized meeting.
She is grateful to the Academy of Finland
(project numbers 163394 and 48787) for financial support.


\begin{thebibliography}{99}

\bibitem{rs} L. Randall, R. Sundrum, \Journal{\PRL}{83}{3370}{1999};
 \Journal{\PRL}{83}{4690}{1999}.

\bibitem{chy1} M. Chaichian, K. Huitu, Z.H. Yu, \Journal{\PLB}{490}{87}{2000}.

\bibitem{chy2} M. Chaichian, K. Huitu, Z.H. Yu, \Journal{\PLB}{508}{317}{2001}.

\bibitem{chky} M. Chaichian, K. Huitu, A.B. Kobakhidze, Z.H. Yu,
hep-ph/0106077, to appear in {\PLB}.

\bibitem{add} B.C. Allanach, A. Dedes, H.K. Dreiner,
\Journal{\PRD}{60}{075014}{1999}.

\bibitem{raby} S. Raby, \Journal{\PRD}{56}{2852}{1997},
{\PLB}{422}{158}{1998};
S. Raby, K. Tobe, \Journal{\NPB}{539}{3}{1999};
A. Mafi, S. Raby, \Journal{\PRD}{62}{035003}{2000}.

\bibitem{fop} A. Faroggi, K. Olive, M. Pospelov, {\it
Astropart. Phys.} {\bf 13}, 31 (2000).

\bibitem{bartl} A. Bartl et al, \Journal{\NPB}{502}{19}{1997}

\bibitem{chry} M. Chaichian, K. Huitu, S. Roy, Z.H. Yu,
hep-ph/0107111.

\bibitem{gw} W.D. Goldberger, M.B. Wise,
\Journal{\PRL}{83}{4922}{1999};\Journal{\PRD}{60}{107505}{1999};
M. Luty, R. Sundrum, \Journal{\PRD}{62}{035008}{2000}.

\bibitem{grw} G. Giudice, R. Rattazzi, J.D. Wells,
\Journal{\NPB}{595}{250}{2001}.

\end{thebibliography}
\end{document}